\documentclass[iop]{emulateapj-rtx4} 
\usepackage{amsmath,natbib,graphicx}
\usepackage{epsf}
\input{epsf}
\usepackage{epstopdf}
\usepackage{multirow}
\usepackage{epsfig}
\usepackage{color}
\usepackage{lscape}
\usepackage{epsfig}
\DeclareGraphicsExtensions{.jpg,.pdf,.png,.eps,.ps}

\newcommand{\be}{\begin{equation}}
\newcommand{\ee}{\end{equation}}
\newcommand{\bea}{\begin{eqnarray}}
\newcommand{\eea}{\end{eqnarray}}

\newcommand{\ltsima}{$\; \buildrel < \over \sim \;$}
\newcommand{\ltsim}{\lower.5ex\hbox{\ltsima}}

\begin{document}

\title{Using the Bullet Cluster as a Gravitational Telescope to Study $z\gtrsim7$ Lyman Break Galaxies}

\author{Nicholas Hall\altaffilmark{1}}
\author{Maru\v{s}a Brada\v{c}\altaffilmark{1}}
\author{Anthony \ H. Gonzalez\altaffilmark{2}}
\author{Tommaso Treu\altaffilmark{3,x}}
\author{Douglas \ Clowe\altaffilmark{4,y}}
\author{Christine\ Jones\altaffilmark{5}}
\author{Massimo Stiavelli\altaffilmark{6}}
\author{Dennis Zaritsky\altaffilmark{7}}
\author{Jean-Gabriel Cuby\altaffilmark{8}}
\author{Benjamin Cl\'ement\altaffilmark{8}}

\shortauthors{Hall et al.}
\altaffiltext{1}{Department of Physics, University of California, One Shields Avenue, Davis, CA 95616}
\altaffiltext{2}{Department of Astronomy, University of Florida, 211 Bryant Space Science Center, Gainesville, FL 32611, USA}
\altaffiltext{3}{Department of Physics, University of California, Santa Barbara, CA 93106, USA}
\altaffiltext{4}{Department of Physics \& Astronomy, Ohio University, Clippinger Labs 251B, Athens, OH 45701
}
\altaffiltext{5}{Harvard-Smithsonian Center for Astrophysics, 60 Garden Street, Cambridge, MA 02138, USA}
\altaffiltext{6}{Space Telescope Science Institute, 3700 San Martin Drive, Baltimore, MD 21218, USA}
\altaffiltext{7}{Steward Observatory, University of Arizona, 933 N Cherry Ave., Tucson, AZ 85721, USA}
\altaffiltext{8}{Laboratoire d'Astrophysique de Marseille, OAMP, Universit\'e Aix-Marseille \& CNRS, 38 rue Fr\'ed\'eric Joliot Curie, 13388 Marseille cedex 13, France}

\altaffiltext{x}{Sloan Fellow, Packard Fellow}
\altaffiltext{y}{Sloan Fellow}
\email{nrhall@ucdavis.edu}

\begin{abstract}
We use imaging obtained with the Hubble Space Telescope Wide Field Camera 3 to search for z$_{850}$ dropouts at $z\sim7$ and J$_{110}$ dropouts at $z\sim9$ lensed by the Bullet Cluster.  In total we find 10 z$_{850}$ dropouts in our $8.27~\rm{arcmin}^2$ field.  Using magnification maps from a combined weak and strong lensing mass reconstruction of the Bullet Cluster and correcting for estimated completeness levels, we calculate the surface density and luminosity function of our z$_{850}$ dropouts as a function of intrinsic (accounting for magnification) magnitude.  We find results consistent with published blank field surveys, despite using much shallower data, and demonstrate the effectiveness of cluster surveys in the search for $z\sim7$ galaxies.  
\end{abstract}


\bigskip\bigskip

\section{Introduction}
\label{sec:intro}

The properties of the first galaxies in the universe are of particular interest given their likely role in the reionization of the intergalactic medium (IGM).  This phase change of the universe from highly neutral to highly ionized is believed to have begun several hundred million years after the Big Bang and ended $\lesssim900$ million years after the Big Bang (corresponding to $z\gtrsim6$) (see e.g., \citealp{fan06b,larson10}).  However, recent observations of $z\gtrsim6$ star-forming galaxies are not yet conclusive as to whether the observed bright end of the population could alone be responsible for reionization, or whether we need to invoke faint objects in higher numbers or with larger stellar masses producing more rest-frame UV photons (e.g., \citealp{bouwens08,stark09,bouwens11}).  To answer this question, more high redshift observations that probe the faint end of the population are essential.  Observational constraints on the fraction of photons from these early galaxies that escape into the IGM, and the clumpiness of the neutral Hydrogen (HI) in the IGM are also needed.

Lyman break galaxies (LBGs) provide the largest and best-studied sample of high redshift galaxies (see e.g., \citealp{vanzella09} for a recent study and \citealp{giavalisco02} for a review).  Their spectra exhibit a sharp decrease in energy at wavelengths below the Lyman limit (the Lyman Break) due to more energetic photons being absorbed within the LBG itself.  For LBGs at $z\gtrsim6$, further absorption below Ly$\alpha$ (the Ly$\alpha$ Break) occurs in intervening HI clouds.  Their high rest-frame UV luminosity blueward of Ly$\alpha$ and distinctive spectral break enable detection of these galaxies out to high redshift through the ``dropout" technique.  This technique utilizes the significant drop in galaxy flux as you move blueward of the break.  LBGs have been detected at $z\sim5$ as V-band dropouts, $z\sim6$ as i-band dropouts,  $z\sim7$ as z-band dropouts, $z\sim8$ as Y-band dropouts, and even $z\sim9-10$ as J-band dropouts. (e.g., \citealp{giavalisco04a,henry08,stark09,castellano10,bouwens11}).  Standard selection criteria have been developed for detecting dropouts in a desired redshift range and avoiding interlopers outside that range (e.g., \citealp{giavalisco04a,beckwith06,bouwens07,bouwens08}).  As an example of their effectiveness, \citet{vanzella09} using followup spectroscopic observations find that only 10\% of their $z\sim4$, 5 and 6 dropout-selected sources from the Great Observatories Origins Deep Survey (GOODS) are interlopers (83\% of which are galactic stars).

Despite the effectiveness of the dropout technique, high redshift galaxy detection is greatly complicated not only by low apparent brightness due to high luminosity distances but also low intrinsic brightness due to low stellar masses compared to moderate redshift ($z\sim2-3$) galaxies (e.g., \citealp{bouwens07}).  This results in faint observed magnitudes for sources at $z\sim7$.  The ability to detect high redshift galaxies can be greatly enhanced by the use of galaxy clusters as gravitational telescopes, as proposed by \citet{soucail90}, in that these galaxies are magnified and thus have brighter observed magnitudes.  This method has been implemented successfully by various authors to study galaxies over a wide range of redshifts (e.g., \citealp{altieri99,blain99,ellis01,metcalfe03,richard06,richard08,bouwens09,zheng09,bradley11}) and is responsible for some of the highest redshift galaxies detected at the time of their discovery (e.g., \citealp{kneib04,bradley08}).

The light from distant galaxies  can be magnified by several orders of magnitude as a result of the large gravitational potential well of an intervening massive galaxy cluster.
This magnification provides increased depth at the cost of decreased observed solid angle.  By way of number counts, the effective slope of the Luminosity Function (LF) being sufficiently steep ($\beta_{eff}=-d(\log\Phi)/d\log L>2$) over the range of magnitudes probed results in a positive magnification bias in that reaching fainter magnitudes by means of lensing more than compensates for the fewer sources resulting from decreased observational solid angle (e.g., \citealp{broadhurst95}). 
Such steep effective slopes have been found at $z\sim7$ all the way down to $L\sim 0.1L^\ast$ (or M$_{1600\AA}\simeq-17.8$, \citealp{bouwens11})\footnote{For the LF parameterized following \citealp{schechter76}, where $\beta_{eff}=-\alpha + L/L^\ast$)}.  Going to even fainter magnitudes can still result in increased cumulative number counts as the loss at low luminosity is overcompensated by the steepness of the LF at the bright end.  Thus the surface density of LBGs can increase significantly as a result of the magnification provided by cluster surveys yielding a much improved statistical analysis of the numerous fainter sources.

In addition to the number of galaxies detected, magnification has the positive consequence of increased spatial resolution, thus allowing better examination of high redshift galaxies on smaller physical scales.  In blank field surveys of color-selected galaxies it is often impossible to confirm the nature of candidate high redshift galaxies given the similarity of their colors with cold Milky Way stars.  Also, due to photometric scatter lower redshift ($z\sim1-2$) elliptical galaxies and moderate redshift ($z\sim2-3$) dust-reddened star-forming galaxies can enter the high redshift selection window.  In cluster surveys however lensed galaxies can be elongated, if they are resolved, thus removing stars as contaminants.  Clusters can also multiply-image background galaxies in a redshift-dependent way, thus removing lower redshift galaxy contaminants.  For multiply-imaged systems, the positions of the images are redshift dependent. A $z>5$ multiply-imaged source can therefore be readily distinguished from a multiply/singly imaged $z\sim 2$ source and foreground (single) stars, provided accurate lensing maps are available.  Cluster surveys are therefore better capable than blank field surveys of differentiating high redshift galaxies from interlopers.  In addition, the magnification provided by clusters can better facilitate spectroscopy of $z\gtrsim7$ galaxies, like that of \citealp{vanzella11} in which the first robust LBGs were confirmed at $z>7$ (see also, \citealp{santos04,stark07b}).

There are various ingredients that make a specific cluster a better suited cosmic telescope for the study of such high-z galaxies.  A more massive cluster produces larger areas of intermediate to high ($\mu>2$) magnification.  A low redshift cosmic telescope would be desirable (as the angular area of high magnification is large), however this results in undesirable amounts of obscuration of background galaxies by cluster members.  Hence clusters of intermediate ($\sim0.3-0.5$) redshift are better suited for these studies.  Detailed cluster magnification maps are essential for the lensed galaxy analysis.  These maps are best constrained in clusters with many strongly lensed images.  In addition, lenses that are highly elliptical in the plane of the sky are desirable because they have larger areas of high magnification.  As a result they also tend to have more arcs and are thus easier to model.  Given these considerations, cluster 1E0657-56 (Bullet Cluster) is in many ways ideal.  Discovered by \citet{tucker95}, it is very hot and X-ray luminous, highly elongated, and at an optimal redshift for these studies ($z\sim0.296$).

In this work we search for $z\gtrsim7$ galaxies in new deep infrared (IR) data of the Bullet Cluster obtained with the Wide Field Camera 3 (WFC3, \citealp{kimble08}) aboard the Hubble Space Telescope (HST).  Following \citet{bradac09}, we optimally combine strong lensing data (including the positions of the multiply-imaged source discussed below) with ground and space-based weak-lensing data using adaptive grid to reconstruct the magnification map of the Bullet Cluster.  With this map we estimate the intrinsic brightness of the LBGs that we detect as dropouts as well as the actual observed solid angle of our observations.

This paper is organized as follows.  We describe the observations and data products in Section~\ref{sec:observations}.  In Section~\ref{sec:reconstruct} we discuss our magnification maps of the Bullet Cluster.  In Section~\ref{sec:photometry} we discuss our photometry.  In Section~\ref{sec:selection} we outline the selection criteria employed to search for $z\gtrsim7$ galaxies, consider various possible sources of contamination, and present our final dropout sample.  We estimate the completeness of our search in Section~\ref{sec:complete}.  In Section~\ref{sec:counts} we consider the surface density and luminosity function of our dropout sample.  We summarize main results in Section~\ref{sec:conclusions}. Where necessary, we assume cosmological parameters consistent with 7-year {\it Wilkinson Microwave Anisotropy Probe} (WMAP) data: $\Omega_m=0.26$, $\Omega_\lambda=0.74$, $h=0.71$, and $\sigma_8=0.8$.  All the coordinates in this paper are given for the epoch J2000.0.  All magnitudes are given in the AB system \citep{oke83}.

\section{Observations and Data Reduction}
\label{sec:observations}

We obtained HST/WFC3 imaging on 19-20 November 2009 (Cycle 16, proposal 11099, PI Brada\v{c}), and with the Advanced Camera for Surveys (ACS, \citealp{ford03}) on 12-13 October 2006 (Cycle 15, proposal 10863, PI Gonzalez) and 12 October 2004 (Cycle 13, proposal 10200, PI Jones). The new WFC3 data consist of two WFC3/IR pointings in F110W (hereafter J$_{110}$, 6530s per pointing) and F160W (hereafter H$_{160}$, 7030s).  The two pointings have an overlapping region centered on the cluster in which the above quoted exposure times are doubled, which are the numbers quoted in \citet{gonzalez10}.  The ACS/WFC imaging used here consists of two pointings centered on the main cluster. The main cluster has been imaged in F606W (hereafter V$_{606}$, 2340s), F775W (hereafter i$_{775}$, 10150s), and F850LP (hereafter z$_{850}$, 12700s), while the subcluster has been imaged in F435W(2420s), F606W (2340s), and F814W (7280s). For the high-redshift galaxy survey described below, we primarily use the main cluster's V$_{606}$, i$_{775}$, z$_{850}$, J$_{110}$, and H$_{160}$ data as described below.

We use the Multidrizzle \citep{koekemoer02} routine to align and combine the images. To register the images we determine the offsets among the images by extracting high S/N objects in the individual, distortion corrected exposures. We use SExtractor \citep{bertin96} and the IRAF routine geomap to identify the objects and calculate the residual shifts and rotations of individual exposures, which were then fed back into Multidrizzle. We use square as the final drizzling kernel and an output pixel scale of $0.1\arcsec$, smaller than the original pixel scale of the WFC3/IR CCD allowing us to exploit the dithering of the observations and improve the sampling of the point-spread function (PSF).  Limiting magnitudes were estimated using $0.63\arcsec \times 0.63\arcsec$ square apertures, which is approximately the same area on average that SExtractor attributed to each object (obtained from the segmentation file) in our final z$_{850}$ dropout catalog. The $5\sigma$ limiting magnitudes for V$_{606}$, i$_{775}$, z$_{850}$, J$_{110}$, and H$_{160}$ are 27.0, 26.4, 26.0, 26.4, and 26.1, respectively.

We also utilize data from the Infrared Array Camera (IRAC, \citealp{fazio04}) onboard NASAs Spitzer Space Telescope ({\it Spitzer}) to assist in the discrimination of contaminants in our high redshift LBG search.  These data were obtained on 2004 December 17-18 and include imaging in all four IRAC bands ([3.6 $\mu$m], [4.5 $\mu$m], [5.8 $\mu$m], and [8 $\mu$m]) with effective exposure times of 4 ks in each filter (see \citealp{gonzalez09} for details).

\section{Magnification Maps of the Bullet Cluster}
\label{sec:reconstruct}

An accurate mass distribution of the Bullet Cluster is needed to obtain a magnification map from which the intrinsic (true, accounting for magnification) magnitudes of our high redshift dropouts can be determined given their lensed (observed) magnitudes.  This mass distribution is reconstructed from the optimal use of multiply-imaged systems (strong lensing) and distortions of singly imaged background sources (weak lensing).  The reconstruction is described in detail in \citet{bradac09} (for some other examples of reconstruction techniques using strong and weak lensing see \citealp{natarajan96,kneib03,smith05,diego07, jee07, limousin07, merten09}).  The reconstruction does not assume a specific form of the underlying gravitational potential.  It is performed on a pixelized adaptive grid, with more pixels nearer to the cluster and multiple images where the S/N of lensing is higher.  Specifically, the course grid (in the outskirts where only weak-lensing data is available) is set to be $18 \arcsec/{\rm pixel}$. The grid is refined by a factor of 4 in the inner circular region (1.5\arcmin\, radius around the Brightest Cluster Galaxy), and by a factor of 16 in the regions where we see multiple images. This gives a final resolution of $\sim1\arcsec/{\rm pixel}$.  For the reconstruction we use the ACS and WFC3 images described above as well as the data used in \citet{bradac06} and \citet{clowe06}.  We obtain a new model closely following the methodology of \citet{bradac09}.  In this model we use the catalog of strongly lensed images from \citet{bradac09}, with the addition of the spectroscopically measured redshift of multiply-imaged system K (z=2.79; \citealp{gonzalez10}).  We also experiment with including the potential multiply-imaged system discussed in Section~\ref{sec:final}.  Including this system results in a change in our intrinsic counts that is well within the error bars.  The weak-lensing catalogs are taken from \citet{clowe06} and are obtained from a combination of ACS and ground-based data that extend to a large FOV; we use the inner $9\arcmin\times9\arcmin$ here. 

Small-scale substructure not accounted for in the lens model (due to a finite number of multiply-imaged systems and erroneous redshifts) will result in uncertainties in the magnification maps of our field.  These magnification uncertainties will produce errors in our intrinsic (in absence of lensing) magnitudes and solid angle.  Magnification error ($\sigma_\mu$) maps from \citet{bradac09} conservatively model the effects of this substructure as follows.  The main cluster and sub clusters were modeled as pseudoisothermal elliptical mass densities and the cluster galaxies were modeled as 30 Singular Isothermal Spheres (SISs).  50 mass clumps containing 8\% of the total mass, which changes critical curve positions by $5-10\arcsec$ (more than the uncertainty with which multiple images are recovered), were also randomly distributed throughout the mass map.  100 realizations of this cluster substructure were generated.  The errors on the magnification were calculated as the standard deviation of the mean magnification at each pixel.  While we produce a new $z=7$ magnification map, which includes the candidate multiple image of this work, we do not produce a new $\sigma_\mu$ map.  Instead we use the map from \citet{bradac09} for a source at $z=6$.  This will be practically indistinguishable from the $\sigma_\mu$ map of a source at $z\sim7$ due to the fact that the lensing strength (or angular diameter distance ratio between lens and source, and observer and source) at these redshifts for a lens at $z\sim0.3$ is nearly constant (see e.g., \citealp{Bartelmann01}).

Given that our $\sigma_\mu$ map used a previous mass reconstruction we cannot use the $\sigma_\mu$ values at the position of the dropouts.  We also cannot use the average $\sigma_\mu$ at the magnification of the dropout, because the actual $\sigma_\mu$ may differ significantly from the average depending on position in the image plane.  We therefore estimate $\sigma_\mu$ at the position of each dropout by using the $\sigma_\mu$ from the map at the pixel nearest the dropout where the mean magnification (from the 100 substructure realizations above) is within 10\% of the magnification of the dropout.  For our dropout sample the average distance from the dropout to this pixel is $10\arcsec$.  These are the magnification errors that appear in Table~\ref{tab:drops} and that are also included in the error on the intrinsic magnitudes by adding them in quadrature with the photometric errors.  

Errors in the magnitudes (lensed and intrinsic) will give rise to errors in the number of dropouts in each magnitude bin of the surface density and LF of our dropout sample.  As an example, there may be a magnitude bin that contains no counts but does contain a significant fraction of the error bars of the dropouts that fall in neighboring bins.  We correct our surface density and luminosity function for this by using the sum of the fractions of the 1$\sigma$ error bars as the number of objects in each bin.

\section{Photometry}
\label{sec:photometry}
Galaxies were detected in the HST/WFC3 imaging data described in Section~\ref{sec:observations}.  We followed a procedure very similar to that of \citet{stark09} and \citet{coe06}. We use SExtractor \citep{bertin96} in dual image mode with detection performed for our z$_{850}$ dropout search in a combined $\rm{J_{110}}+\rm{H_{160}}$ image, thus improving the signal to noise and the measurement of the centroid, and for our J$_{110}$ dropout search in an H$_{160}$ image.  Photometry was performed in each of the V$_{606}$, i$_{775}$, z$_{850}$, J$_{110}$, H$_{160}$ filters.
All magnitudes were first corrected for foreground galactic extinction given the B-V color excess in the direction of the bullet cluster \citep{schlegel98}, and the corresponding extinction values from \citet{sirianni05}\footnote{Assuming an 05 V type star since these early star-forming galaxies should be dominated by young massive stars} for V$_{606}$, i$_{775}$, and z$_{850}$ and from \citet{schlegel98} for J$_{110}$ and H$_{160}$.  
This yields foreground galactic extinction values of (0.23, 0.16, 0.12, 0.07, 0.05) for (V$_{606}$, i$_{775}$, z$_{850}$, J$_{110}$, H$_{160}$), respectively.  Aperture magnitudes were measured in fixed circular apertures of 0.8\arcsec~ diameter with SExtractor's MAG\_APER and were used to measure colors.  We correct the aperture magnitudes for varying PSF size as described below.  The final magnitudes in each filter were computed using a combination of aperture colors and the SExtractor parameter MAG\_AUTO of the reddest filter (H$_{160}$), i.e., 
\bea
\rm{m_x} &=& {\rm MAG\_AUTO\left(H_{160}\right) + } \nonumber \\
&& \rm{\left(MAG\_APER\left(H_{160}\right) - MAG\_APER\left(x\right)\right),}
\eea
where x is V$_{606}$, i$_{775}$, z$_{850}$, and J$_{110}$.

\begin{figure*}[ht]\centering
\includegraphics[scale=0.51]{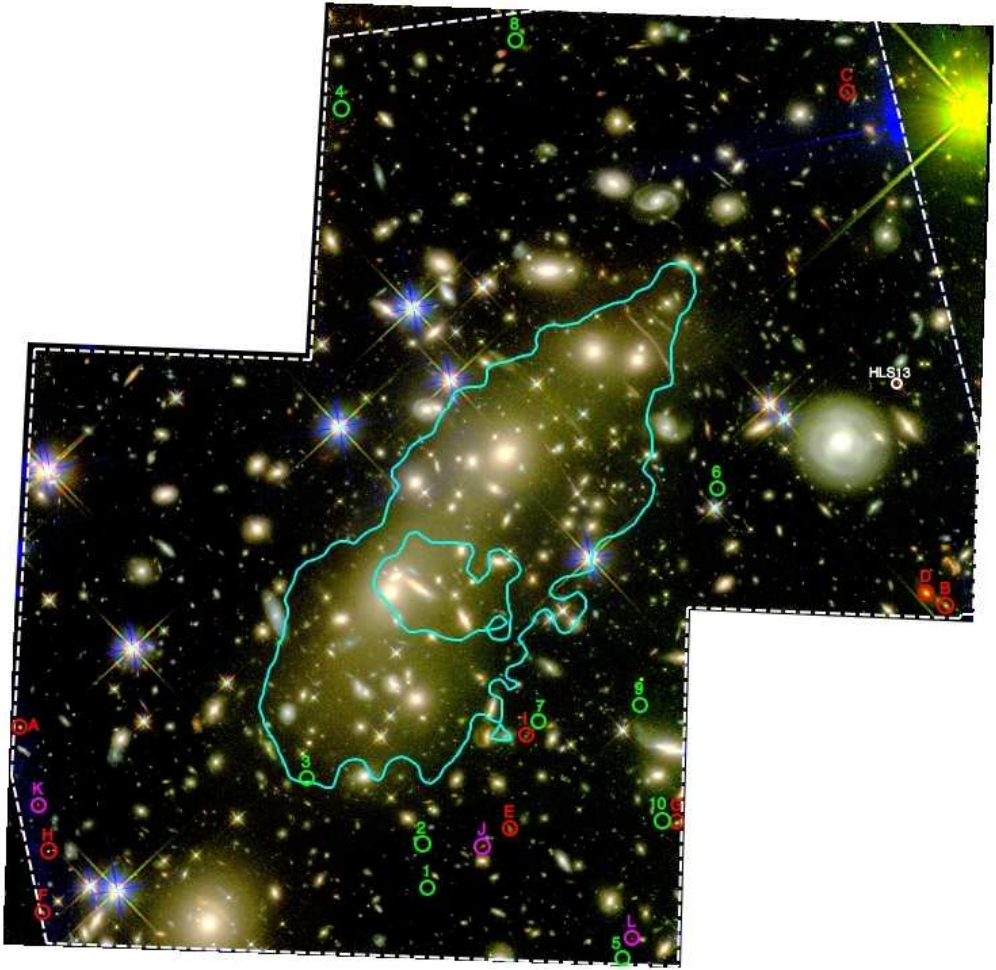}
 \caption[]{\small A combined color image of the Bullet Cluster (z$_{850}=$ blue image, J$_{110}=$ green image, and H$_{160}=$ red image) with the search area outlined ({\it dashed white line}, showing the tilt of the z$_{850}$ pointing relative to the two WFC3 pointings) and the curve of maximum magnification of the lens model overlaid ({\it cyan line}).  The z$_{850}$ dropout candidates are shown as {\it green circles} labelled with the numbers of Table~\ref{tab:drops}.  The z$_{850}$ (J$_{110}$) dropout contaminants are shown as {\it red {\rm(}magenta{\rm)} circles} labelled with the letters of Table~\ref{tab:contaminants}.  The {\it Herschel} source HLS13 of \citet{rex10} is shown as a {\it white circle}.  The image is $3.3\arcmin \times 3.3\arcmin$.  North is up and East is left.}
\label{fig:bullet}
\end{figure*}

There are various factors that need to be considered in our photometric analysis.  First, intracluster light (ICL) occupies $\sim20-25\%$ of our search area in J$_{110}$ and H$_{160}$ where it is brightest, as can be seen in the combined color image of Figure~\ref{fig:bullet}.  Difficulties in determining the background given the presence of the ICL and in distinguishing between ICL and galaxy flux may introduce additional errors and/or a bias in our measured magnitudes.  We carefully examine the uncertainties introduced by ICL as described below.  Second, the inability of MAG\_AUTO to pick up the wings of the PSF can lead to measured magnitudes that are too faint.  Finally, the varying size of the PSF for the different wavelengths probed by the filters must be accounted for in the aperture magnitudes.  We investigate our photometric errors and possible biases with detailed simulations in each band as follows.

We first choose objects from the stellar locus and estimate the full-width-half-max (FWHM) of the PSF, which we measure to be (0.16, 0.17, 0.17, 0.18, 0.19) arcsec in V$_{606}$, i$_{775}$, z$_{850}$, J$_{110}$, and H$_{160}$, respectively.  This is slightly larger than the estimates of \citet{bouwens11} for example.  This is expected given that we have fewer images and thus a larger multidrizzle dropsize (1.0), which is essentially the size of the top hat with which the PSF is convolved.  
We place 26th magnitude\footnote{Chosen given that the average H$_{160}$ magnitude of our z$_{850}$ dropouts is 25.9} stars in the image convolved with a PSF generated with Tiny Tim \citep{hook08} with the above measured FHWM.  Regarding the number of stars, we place a sufficient number uniformly in the image so that, given our completeness levels of Section~\ref{sec:complete}, we detect $\sim100$ of them and measure their magnitudes.
We do this 5 times, combine these $\sim 500$ objects, and measure the mean and standard deviation of the distribution of measured magnitude minus input magnitude.  We perform 3 sigma clipping from the median of the distribution to better measure the mean of the magnitude difference by excluding outliers, such as mock objects confused with real objects in the image.  For MAG\_AUTO in H$_{160}$ we measure a mean of 0.15 and standard deviation of 0.16.  For comparison, using a Gaussian PSF we find a mean of -0.02 and standard deviation of 0.18, demonstrating the effect of an extended PSF.  We also do this for the aperture magnitudes used for our colors.  With the Tiny Tim PSF we find a mean of (0.04, 0.06, 0.04, 0.19, 0.18) and standard deviation of (0.11, 0.11, 0.14, 0.09, 0.12) in V$_{606}$, i$_{775}$, z$_{850}$, J$_{110}$, and H$_{160}$, respectively.  Using a Gaussian PSF we find a mean of (0.00, -0.01, -0.01, -0.01, -0.01) and standard deviation of (0.10, 0.11, 0.13, 0.08, 0.12) in V$_{606}$, i$_{775}$, z$_{850}$, J$_{110}$, and H$_{160}$, respectively.  As expected, the bias is smaller at shorter wavelengths for the Tiny Tim PSF due to the smaller PSF size resulting in a larger fraction of the object flux within the $0.8\arcsec$ diameter aperture.  The difference between the ACS and WFC3 PSFs is also manifested.

For both MAG\_AUTO in H$_{160}$ and the aperture magnitudes in all of the bands, we correct for the biases found with the Tiny Tim PSF, which is currently our best estimate of the PSF shape.  Even though the bias depends strongly on the PSF shape, a misestimate of 0.1-0.2 magnitudes would not significantly effect our our surface density and luminosity function given the large bin sizes used (see Sec.~\ref{sec:counts}).  We also use the standard deviations found with the Tiny Tim PSF, which does not depend significantly on the PSF shape, as the minimum photometric error on the magnitudes.  

\section{Dropout Selection}
\label{sec:selection}

In this section we discuss our selection of z$_{850}$ and J$_{110}$ dropouts through the use of color and S/N criteria  (Section~\ref{sec:colsn}), intrinsic brightness cuts (Section~\ref{sec:intrbright}), and infrared ({\it Spitzer} and HAWK-I) data (Section~\ref{sec:infrared}).  We then provide the properties of our final dropout sample as well as potentially interesting contaminants (Section~\ref{sec:final}).  Finally we address the possibility of contamination from point sources.

\subsection{Color and S/N Criteria}
\label{sec:colsn}

We select our z$_{850}$ dropouts by requiring all of the following selection criteria:
\be
\rm{z}_{850}-J_{110} > 1.0 + 0.4(J_{110}-H_{160})
\label{eqn:zdropcol1}
\ee
\be
\rm{z}_{850}-J_{110} > 1.0
\label{eqn:zdropcol2}
\ee
\be
J_{110}-H_{160} \leq 1.1
\label{eqn:zdropcol3}
\ee  
\be
S/N\left(H_{160}\right)>3.5,\ S/N\left(J_{110}\right)>3.5
\label{eqn:s2nhj}
\ee
\be
S/N\left(i_{775}\right)<3,\ S/N\left(V_{606}\right)<3.
\label{eqn:s2niv}
\ee
We will refer to Equations \ref{eqn:zdropcol1}, \ref{eqn:zdropcol2}, and ~\ref{eqn:zdropcol3} as the  ``color criteria''  and Equations \ref{eqn:s2nhj} and \ref{eqn:s2niv} as the ``S/N criteria".
The color criteria are adapted from those of \citet{bouwens08}, which were developed for HST/NIC3 J$_{110}$ and H$_{160}$.  To adjust them to be consistent with our filters we shift them by the amount by which the colors differ for a $z=7$ template star-forming galaxy spectrum when measured with our filters versus those of \citet{bouwens08} (a difference of $\sim0.2$ magnitudes).  The template spectrum we use is from the isochrone synthesis code of \citet{bruzual03} with an instantaneous burst, an age of 100 Myr, metallicity of $Z=0.008 Z_\sun$, and no dust (See e.g., \citealp{labbe10,castellano10b}).  We refer to this template spectrum below as our ``BC03 starburst template".

The S/N cut of Equation~\ref{eqn:s2nhj} corresponds to the requirement S/N$\left(J_{110}+H_{160}\right)\gtrsim5\sigma$.  Deep blank field surveys typically impose S/N cuts of $2\sigma$ for the filters blue-ward of the dropout filter.  However, given that we have less exposures to combine, we are likely to have more artifacts due to cosmic rays that can cause artificial detections.  We thus impose $3\sigma$ S/N\footnote{Estimated as the flux divided by the flux error using the photometry discussed in Section~\ref{sec:photometry}.} cuts and then carefully inspect each object by eye in the V$_{606}$ and i$_{775}$ images and throw it out if it is detected.  We also remove artifacts such as diffraction spikes and fake objects near the detector edges. 

\begin{figure}[ht]\centering
\includegraphics[scale=0.7]{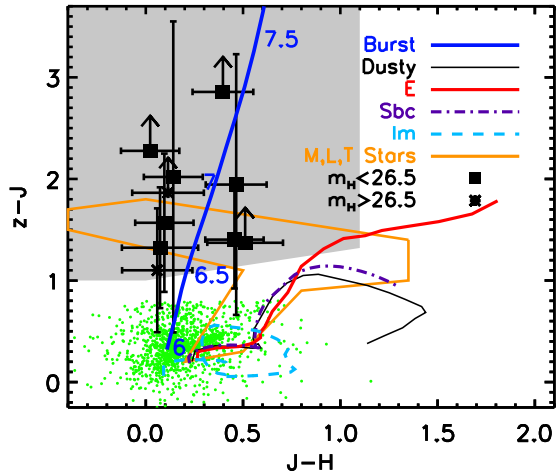}
 \caption[]{\small z$_{850}$-J$_{110}$ versus J$_{110}$-H$_{160}$ color-color diagram showing our z$_{850}$ dropout selection window ({\it grey shaded region}).  The colors of our z$_{850}$ dropouts with $\rm{m}_{H_{160}}<26.5$, those used in the surface density and luminosity function, are shown ({\it black squares}) as well as those with $\rm{m}_{H_{160}}>26.5$ ({\it black asterisks}).  Sources which do not satisfy our dropout selection criteria are shown in green.  The redshift evolution of the colors of a \citet{bruzual03} star-forming galaxy with low metallicity ($Z=0.008Z_\sun$), an age of 100 Myr, no dust, and attenuated for IGM absorption as a function of redshift according to \citet{madau95} is shown with a {\it blue line}.  The colors  of a \citet{chary01} dusty galaxy (Dusty,{\it thin black line}) and a \citet{coleman80} elliptical (E, {\it red line}), spiral (Sbc, {\it dot-dashed purple line}), and irregular galaxy (Im, {\it dashed cyan line}) are also shown as they evolve in redshift.  The {\it orange lines} show region in which we expect T Dwarf colors to lie \citep{knapp04}.
 }
\label{fig:color}
\end{figure}

The above z$_{850}$ dropout selection criteria select galaxies in the redshift range $6 \lesssim z \lesssim 9$ with a mean redshift of 7.5 for objects with $\rm{m}_{H_{160}}=26$ (see Figure~\ref{fig:redshift} of Section~\ref{sec:complete}).   The selection window is shown as the grey shaded region in the z$_{850}$-J$_{110}$ versus J$_{110}$-H$_{160}$ color Figure~\ref{fig:color}.  Sources that do not satisfy our dropout selection criteria are shown as green dots.  The colors of low-mass M, L, and T dwarf stars are expected to lie in the region enclosed by orange lines.  The typical colors of various galaxy types are shown as they evolve in redshift (for our BC03 starburst template, a dusty galaxy from \citealp{chary01}, and an elliptical, spiral, and irregular galaxy from \citealp{coleman80}).  

We also search for J$_{110}$ dropouts by requiring all of the following color and S/N selection criteria:
\be
J_{110}-H_{160} > 1.1
\ee  
\be
S/N\left(H_{160}\right)>5
\ee
\be
S/N\left(z_{850}, i_{775}, V_{606}\right)<3.
\ee
Again, we also carefully inspect each object in the images and throw it out if it is detected in V$_{606}$, i$_{775}$, or z$_{850}$.  The above J$_{110}$ dropout selection criteria select galaxies in the redshift range $8 \lesssim z \lesssim 11$ with a mean redshift of 9.5  for objects with $\rm{m}_{H_{160}}=26$ (see Figure~\ref{fig:redshift} of Section~\ref{sec:complete}).  

\subsection{Intrinsic Brightness Cuts}
\label{sec:intrbright}

The $4000\AA$ break of a $z\sim1.4$ (2.0) elliptical galaxy will be at the same wavelength as the Lyman Break for a $z\sim7$ (9) LBG.  These elliptical galaxies as well as dust-reddened galaxies may enter the red portion of our selection window through photometric scatter.

These lower redshift galaxies are on average intrinsically brighter than the high z LBGs we are targeting.  We thus exclude bright interlopers by rejecting those objects intrinsically brighter than the brightest LBG we might expect to see.  We estimate the brightest intrinsic H$_{160}$ apparent magnitude we might see as follows.  A $10L^\ast$ galaxy\footnote{We use the \citet{bouwens11} best fit $z\sim7$ (8) LF to determine the $L^\ast$ used for z$_{850}$ (J$_{110}$) dropout intrinsic brightness cut.} at $z=6$ (7.5), the lowest redshift allowed by our z$_{850}$ (J$_{110}$) selection criteria, would have an apparent magnitude of 24.0 (24.2).  We reject 2 z$_{850}$ dropout candidates and 1 J$_{110}$ dropout candidate that have intrinsic H$_{160}$ apparent magnitudes $\rm{m}_{H_{160,{\rm int}}}$ (obtained from the observed $m_{H_{160}}$ and the magnification at the position of the object from the $z=7$ magnification map) brighter than this cut.

\subsection{Infrared Data}
\label{sec:infrared}

Whereas bright lower redshift elliptical and dusty galaxies can be rejected with S/N and intrinsic magnitude cuts, fainter objects are not as easily excluded.  Contamination from faint elliptical and dusty galaxies in our z$_{850}$ dropout catalog is expected to be minimal given that our candidates do not have colors consistent with them (to at least one sigma, see Figure~\ref{fig:color}).  We nevertheless utilize our longer wavelength data to aid in distinguishing these contaminants from high redshift LBGs.  For our z$_{850}$ dropouts we consider the H$_{160}$ - [$3.6\mu$m] ({\it Spitzer} Chanel 1) color of these galaxies.  Whereas our BC03 starburst template at $z=7$ has H$_{160}$ - [$3.6\mu\rm{m}]=1$, many ellipticals and dusty galaxies are significantly redder than this.  For example, both the elliptical and dusty galaxy templates used in Figure~\ref{fig:color} begin to have H$_{160}$ - [$3.6\mu\rm{m}]>1$ at $z=1.3$.  By $z=3$ the elliptical (dusty) template has H$_{160}$ - [$3.6\mu\rm{m}] = 2.8$ (2.1).  The H$_{160}$ - [$3.6\mu$m] color for dusty galaxies can vary widely depending on how dusty the galaxy is and its redshift.  As an extreme example, we consider the dusty lensed galaxy behind the bullet cluster, which has $A_V\sim3.8$ \citep{gonzalez10}. We can detect a galaxy such as this one, with $H_{160}$ - [$3.6\mu\rm{m}]=3.75$, at $3\sigma$ in [3.6$\mu$m] for sources as faint as $\rm{m}_{H_{160}}=26$.  Whereas, some of these fainter ellipticals and dusty galaxies will be detected in [3.6$\mu$m], we do not expect our z$_{850}$ dropouts to be, which is in fact the case for all of the objects in our sample.

There are two objects that satisfy our J$_{110}$ color criteria, S/N criteria, and intrinsic brightness cut, referred to below as objects J and L in Table~\ref{tab:contaminants} and Figure~\ref{fig:contaminants} (object K does not satisfy the intrinsic brightness cut).  They are both quite bright; object J (L) has $\rm{m}_{H_{160,{\rm int}}}=24.8\pm0.2$ ($24.6\pm0.2$) and is thus only $0.6\pm0.2$ ($0.4\pm0.2$) magnitudes fainter than the intrinsic brightness cut.  Both objects are clearly detected in [4.5$\mu$m] in our {\it Spitzer} data.   In [3.6$\mu$m] object J is clearly detected but object L can't unambiguously be separated from another nearby brighter galaxy.  We do not expect our high redshift LBGs to be detected in [3.6$\mu$m] (as discussed above) or in [4.5$\mu$m].  In addition, both of these objects are detected in the Ks-band of the High Acuity, Wide field K-band Imaging (HAWK-I, \citealp{kissler-patig08}) camera on the Very Large Telescope (VLT) and have H$_{160}$ - Ks $\sim0.5$.  Our BC03 starburst template would have to be at $z\sim11$ in order to have an H-K color 0.5.  It is very unlikely that these objects are at such high redshift given their intrinsic brightnesses.  On the other hand, a dusty galaxy spectrum from \citet{chary01} at $z=1.9$ (2.3) fits the J$_{110}$, H$_{160}$, and Ks-band data of object J (L) quite well.  These facts lead us to believe that these objects are most likely low redshift interlopers and thus we do not consider them high redshift candidates, and do not include them in the rest of our analysis.

\subsection{Final Dropout Sample}
\label{sec:final}

\begin{figure}[ht]\centering
\includegraphics[scale=0.29]{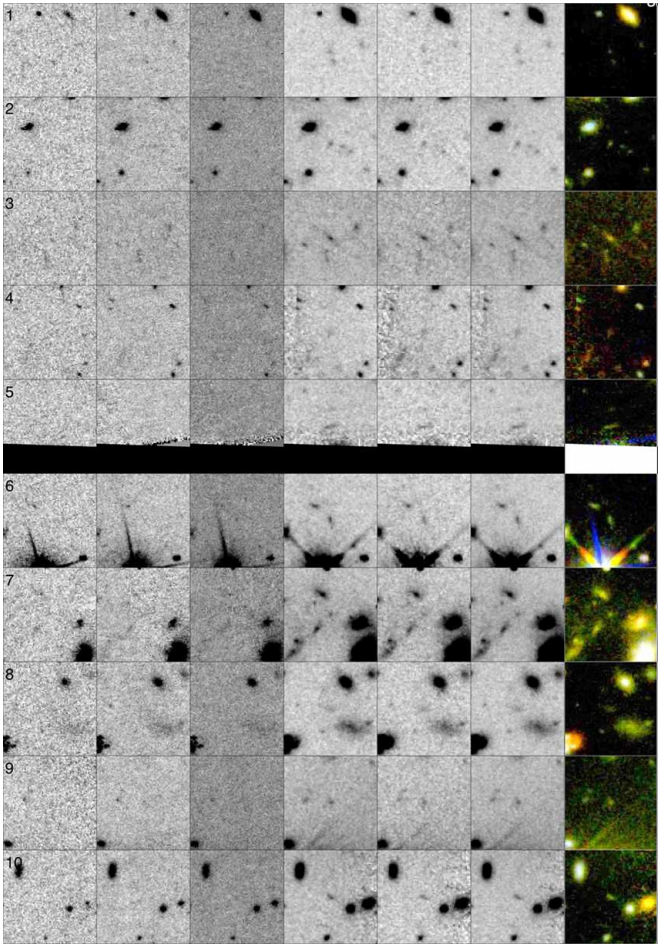}
 \caption[]{\small Cutouts of our z$_{850}$ dropouts shown (from left to right) in V$_{606}$, i$_{775}$, z$_{850}$, J$_{110}$, H$_{160}$, $\rm{J_{110}}+\rm{H_{160}}$, and a z$_{850}$J$_{110}$H$_{160}$ color image.  The numeric labels correspond to the labels of Table~\ref{tab:drops}.  The cutouts are $7\arcsec\times7\arcsec$.  To get an idea of physical size, for a source at $z=7$ (9) and in the absence of lensing, $7\arcsec$ corresponds to 37 (32) kpc.  For a uniform magnification of 12, the highest in our z$_{850}$ dropout sample, $7\arcsec$ corresponds to an intrinsic length of 11 (9.1) kpc for a source at $z=7$ (9).
 }
\label{fig:cuts}
\end{figure}

Upon implementing the above z$_{850}$ dropout selection criteria, we obtain a final sample of 10 z$_{850}$ dropouts, with cutout images shown in Figure~\ref{fig:cuts} and properties given in Table~\ref{tab:drops}.  The z$_{850}$ dropouts are also shown as green circles in Figure~\ref{fig:bullet}.  Only those z$_{850}$ dropouts with $\rm{m}_{H_{160}}<26.5$, where the completeness is greater than $22\%$ (See Section~\ref{sec:complete}) are used in the surface density and luminosity function below.  These 8 objects are shown as the black squares in Figure~\ref{fig:color}, whereas the 2 objects with $\rm{m}_{H_{160}}>26.0$, are shown as asterisks.  Upon implementing all the J$_{110}$ dropout selection criteria, we find no J$_{110}$ dropout candidates.  

\begin{table*}[ht!]
 \begin{center}
 \caption{The Properties of z$_{850}$ Dropouts}
 \small
 \begin{tabular}{cccccccc}
 \hline\hline
\rule[-2mm]{0mm}{6mm}
Dropout & R.A. & Decl. & $\rm{m}_{H_{160}}\pm\Delta_p$ & $\left(J_{110}-H_{160}\right)$ & $\left(z_{850}-J_{110}\right)$ & $\mu$ &$\rm{m}_{H_{160},\rm{int}}\pm\Delta_{l+p}$ \\
\hline
1 & 104.65470 & -55.974464 & 26.77$\pm$0.23 & 0.11$\pm$0.18 & $>$1.3 & 4.3$\pm$0.2 & 28.34$^{+0.24}_{-0.24}$ \\
2 & 104.65527 & -55.971901 & 26.97$\pm$0.23 & 0.06$\pm$0.18 & 1.10$\pm$0.61 & 6.5$\pm$0.5 & 28.99$^{+0.25}_{-0.25}$ \\
3 & 104.66736 & -55.968067 & 24.97$\pm$0.16 & 0.46$\pm$0.15 & 1.40$\pm$0.48 & 12$\pm$4 & 27.67$^{+0.35}_{-0.47}$ \\ 
4 & 104.66375 & -55.928802 & 26.37$\pm$0.22 & 0.51$\pm$0.19 & $>$1.3 & 2.8$\pm$0.08 & 27.46$^{+0.22}_{-0.22}$ \\
5 & 104.63437 & -55.978603 & 25.91$\pm$0.22 & 0.07$\pm$0.20 & 1.32$\pm$0.59 & 2.1$\pm$0.03 & 26.71$^{+0.22}_{-0.22}$ \\
6 & 104.62446 & -55.951065 & 25.85$\pm$0.19 & 0.02$\pm$0.15 & $>$2.1 & 10.$\pm$2 & 28.37$^{+0.28}_{-0.31}$ \\
7 & 104.64304 & -55.964756 & 25.81$\pm$0.26 & 0.14$\pm$0.15 & 2.02$\pm$1.53 & 5.2$\pm$0.5 & 27.61$^{+0.28}_{-0.28}$ \\
8 & 104.64549 & -55.924828 & 25.89$\pm$0.21 & 0.40$\pm$0.16 & $>$1.9 & 3.1$\pm$0.1 & 27.12$^{+0.22}_{-0.22}$ \\
9 & 104.63254 & -55.963764 & 26.00$\pm$0.16 & 0.46$\pm$0.16 & 1.94$\pm$1.28 & 4.3$\pm$0.3 & 27.59$^{+0.17}_{-0.18}$ \\
10 & 104.63015 & -55.970482 & 26.37$\pm$0.16 & 0.10$\pm$0.15 & 1.57$\pm$0.68 & 3.0$\pm$0.2 & 27.57$^{+0.17}_{-0.17}$ \\
\hline
\hline
\label{tab:drops}
\end{tabular}
\tablecomments{The final error on m$_{H_{160,\rm{int}}}$ ($\Delta_{\rm{l+p}}$) includes errors in the lensing (magnification) and photometry added in quadrature.  z$_{850}$-J$_{110}$ lower limits are the 1$\sigma$ limiting magnitudes computed in the vicinity of the dropout.  
}
\normalsize
\end{center}
\end{table*}

The majority of the z$_{850}$ dropouts (7 out of 10) are found in the bottom right portion of Figure~\ref{fig:bullet}.  This is likely due to a combination of small number statistics and the fact that the bottom right portion of the image has a slightly different magnification distribution according to our lens model. There also appears to be a detector artifact in the top left portion of each of the WFC3 pointings of $\sim 250$ arsec$^2$ that is likely impeding detection in those regions.

\begin{figure}[ht]\centering
\includegraphics[scale=0.288]{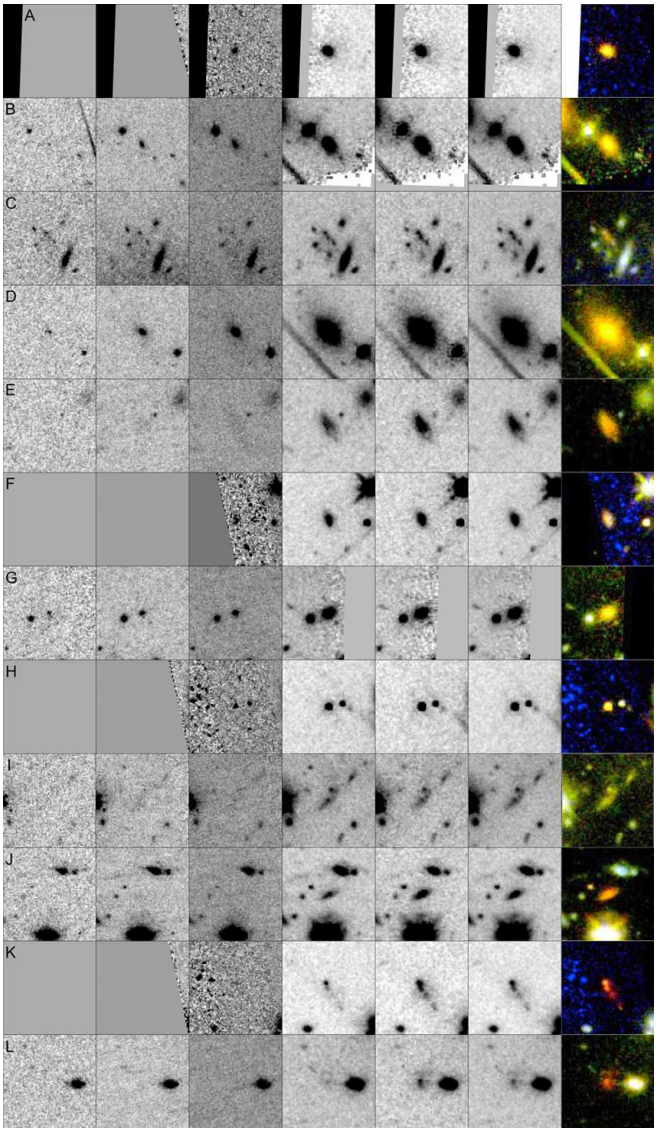}
 \caption[]{\small Cutouts of the partial sample of dropout contaminants of Table~\ref{tab:contaminants}.  These objects have $\rm{m}_{H_{160,int}}<25$ (except object I) and satisfy the z$_{850}$ (top) and J$_{110}$ (bottom) dropout color cuts but do not satisfy the other criteria.  They are shown (from left to right) in V$_{606}$, i$_{775}$, z$_{850}$, J$_{110}$, H$_{160}$, $\rm{J_{110}}+\rm{H_{160}}$, and a z$_{850}$J$_{110}$H$_{160}$ color image.  The labels correspond to the labels of Table~\ref{tab:contaminants}.  The cutouts are $7\arcsec\times7\arcsec$.
 }
\label{fig:contaminants}
\end{figure}

The properties of bright ($\rm{m}_{H_{160,{\rm int}}}<25$) lower redshift objects that satisfy our z$_{850}$ (J$_{110}$) dropout color criteria but are rejected on account of S/N criteria, brightness cuts, or IR ({\it Spitzer}/Hawk-I) data are given in the top (bottom) of Table~\ref{tab:contaminants}.  Cutouts of these objects are also shown in the top (bottom) of Figure~\ref{fig:contaminants} and they are shown as red (magenta) circles in Figure~\ref{fig:bullet}.  We also show the dust-reddened {\it Herschel} source HLS13 of \citet{rex10} as a white circle in Figure~\ref{fig:bullet}.

\begin{table*}[ht!]
 \begin{center}
 \caption{A Partial Sample of z$_{850}$ (top) and J$_{110}$ (bottom) Dropout Contaminants}
 \small
 \begin{tabular}{cccccccc}
 \hline\hline
\rule[-2mm]{0mm}{6mm}
Contaminant & R.A. & Decl. & $\rm{m}_{H_{160}}\pm\Delta_p$ & $\left(J_{110}-H_{160}\right)$ & $\left(z_{850}-J_{110}\right)$ & $\rm{m}_{i_{775}}$ & $\rm{m}_{V_{606}}$ \\
\hline
z$_{850}$ A & 104.69737 & -55.965012 & 21.93$\pm$0.16 & 0.72$\pm$0.15 & 1.34$\pm$0.17 & $>$28.1 & $>$28.7 \\
$\phantom{z_{850}}$ B & 104.60057 & -55.957890 & 21.64$\pm$0.16 & 0.67$\pm$0.15 & 1.51$\pm$0.17 & 24.67$\pm$0.23 & 26.81$\pm$0.37 \\
$\phantom{z_{850}}$ C & 104.61086 & -55.927860 & 23.53$\pm$0.16 & 0.62$\pm$0.15 & 1.62$\pm$0.38 & 25.24$\pm$0.25 & 24.81$\pm$0.24 \\
$\phantom{z_{850}}$ D & 104.60263 & -55.957169 & 20.30$\pm$0.16 & 0.75$\pm$0.15 & 1.46$\pm$0.17 & 23.44$\pm$0.23 & 25.19$\pm$0.23 \\
$\phantom{z_{850}}$ E & 104.64615 & -55.970993 & 22.62$\pm$0.16 & 0.86$\pm$0.15 & 1.47$\pm$0.17 & 25.37$\pm$0.25 & 27.21$\pm$1.07 \\
$\phantom{z_{850}}$ F & 104.69501 & -55.975872 & 23.14$\pm$0.16 & 0.62$\pm$0.15 & 1.25$\pm$0.25 & 26.31$\pm$0.23 & 25.86$\pm$0.23 \\
$\phantom{z_{850}}$ G & 104.62870 & -55.970612 & 21.96$\pm$0.16 & 0.62$\pm$0.15 & 1.38$\pm$0.17 & 24.64$\pm$0.23 & 25.58$\pm$0.23 \\
$\phantom{z_{850}}$ H & 104.69444 & -55.972286 & 22.49$\pm$0.16 & 0.51$\pm$0.15 & 1.53$\pm$0.17 & $>$28.1 & $>$28.7 \\
$\phantom{z_{850}\;}$ I\tablenotemark{*} & 104.64444 & -55.965549 & 24.18$\pm$0.16 & 0.34$\pm$0.15 & 1.36$\pm$0.26 & 25.70$\pm$0.24 & 27.23$\pm$0.73 \\
\hline
J$_{110}$ J & 104.64898 & -55.971981 & 23.12$\pm$0.16 & 1.19$\pm$0.15 & 2.51$\pm$0.48 & 26.78$\pm$0.37 & $>$28.7 \\
$\phantom{J_{110}}$ K & 104.69546 & -55.969612 & 23.10$\pm$0.16 & 1.23$\pm$0.15 & 1.45$\pm$0.63 & $>$28.1 & $>$28.7 \\
$\phantom{J_{110}}$ L & 104.63338 & -55.977394 & 23.69$\pm$0.16 & 1.62$\pm$0.15 & 2.05$\pm$1.08 & 28.16$\pm$1.58 & $>$28.7 \\
\hline
\hline
\label{tab:contaminants}
\end{tabular}
\tablenotetext{*}{This object has $\rm{m}_{H_{160,{\rm int}}}=26.49\pm0.19$. See end of Section~\ref{sec:final}.}
\tablecomments{These objects have $\rm{m}_{H_{160,{\rm int}}}<25$ (except object I) and satisfy our dropout color criteria but are rejected from our sample as low redshift galaxies on account of S/N selection criteria, brightness cuts, or IR ({\it Spitzer}/Hawk-I) data.  Similar photometric errors are due to assigning the minimum error discussed in Section~\ref{sec:photometry}.  Objects A, B, and F lie very near a field edge (see Figure~\ref{fig:bullet}) and thus their photometry may be unreliable.  Lower limits are the 1$\sigma$ limiting magnitudes quoted in Section~\ref{sec:observations}.
}
\normalsize
\end{center}
\end{table*}

Low-mass stars (e.g., L and T Dwarfs), supernovae and active galactic nuclei are possible contaminants that will appear as point sources in our images.  
We take advantage of the high-resolution of WFC3/IR and the increased resolution due to lensing to determine which of our sources are resolved and which are not resolved and thus consistent with point sources.  We do not attempt to estimate the sizes of those objects with $\rm{m}_{H_{160}}>26.5$ given the uncertainty in measuring low surface brightness objects.  Only those objects with $\rm{m}_{H_{160}}<26.5$ are used below in the surface density and LF of Figures~\ref{fig:counts} and~\ref{fig:lf}.  
All of these objects have measured FWHM greater than 0.26\arcsec~in J$_{110}$ and 0.21\arcsec~in H$_{160}$ which is above the PSF size (0.18\arcsec~in J$_{110}$ and 0.19\arcsec~in H$_{160}$).  Thus contamination due to point sources is unimportant.  

Object I of Table~\ref{tab:contaminants} has some characteristics consistent with being a counter-image of object 3 of Table~\ref{tab:drops}.  It satisfies the z$_{850}$ dropout color criteria.  It has $J_{110}-H_{160}$ and $z_{850}-J_{110}$ colors consistent with object 3 within $1\sigma$ error bars.  Both have shapes extended in the direction expected by the lens model for a $z>6$ source, as can be seen in Figure~\ref{fig:bullet} in which the critical curve (cyan line) of the lens model is overlaid and the positions of the two objects are shown.  Finally, given the position of one of the images, the mass model predicts a counter-image in the vicinity of the other (we have also verified that according to the lens model, none of the other dropout candidates would have observable counter-images).

However, this object also has characteristics inconsistent with being a counter-image of object 3.  Its intrinsic H$_{160}$ magnitude of $26.49\pm0.19$ differs significantly from that of object 3.  However, the error on the magnification of object 3 is likely underestimated given that it lies close ($<1\arcsec$) to the critical curve where the magnification changes rapidly with position (magnification errors are expected to be correct on average but may be under or overestimated for individual objects).  Also, it is detected at 4.5$\sigma$ in i$_{775}$ and thus does not satisfy the S/N criterion of Equation~\ref{eqn:s2niv}.

\section{Completeness}
\label{sec:complete}

The completeness was estimated as function of redshift and observed H$_{160}$ mangnitude ($\rm{m}_{H_{160}}$) with simulations using the IRAF task artdata.  For each H$_{160}$ magnitude, $\rm{m}_{H_{160}}=\left(24,25,26,27\right)$, we generated 100 mock galaxies at 25 different equally spaced redshifts in the range $5.5<z<10$.  The colors of each galaxy were determined from a template spectrum, where again we follow \citet{madau95} to apply an intergalactic medium absorption correction to the colors given the redshift.  These galaxies were then added with a spatially uniform distribution to the $\rm{J_{110}}+\rm{H_{160}}$ images as stars convolved with a Tiny Tim PSF scaled to have FWHM of (0.16, 0.17, 0.17, 0.18, 0.19) arcsec in V$_{606}$, i775, z$_{850}$, J$_{110}$, and H$_{160}$, respectively.  We then attempt to detect these galaxies following the same procedure as the actual dropouts, meaning the galaxies must both be detected as well as satisfy the dropout color and S/N selection criteria.  For each $\rm{m}_{H_{160}}$ and $z$, the probability of detecting a galaxy as a dropout $p\left(\rm{m}_{H_{160}},z\right)$, or completeness, is simply the ratio of the number of galaxies detected as dropouts to the number of dropout galaxies added to the images.  This completeness was estimated with five SEDs from the isochrone synthesis code of \citet{bruzual03} with instantaneous bursts and the following age and metallicity (Z in units of $Z_\sun$) combinations: 5 Myr and Z=0.02, 25 Myr and Z=0.02, 100 Myr and Z=0.02, 100 Myr and Z=0.05, 100 Myr and Z=0.008.  For our final completeness we use the average of the completeness estimates from each SED.  

\begin{figure}[ht]\centering
\includegraphics[scale=0.7]{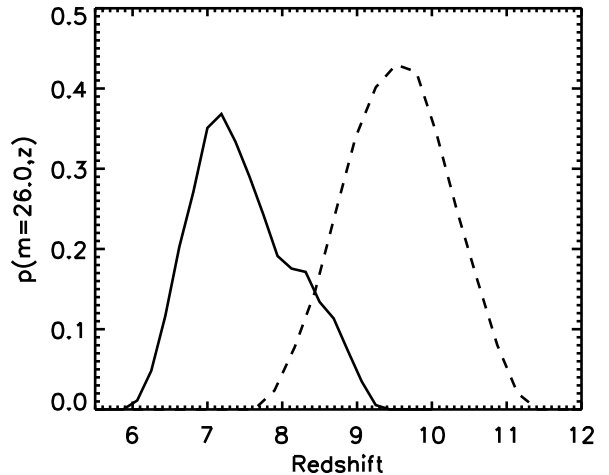}
 \caption[]{\small 
For an LBG with $\rm{m}_{H_{160}}=26$, the probability of both detecting it and it satisfying our color and S/N criteria for z$_{850}$ ({\it solid line}) and J$_{110}$ ({\it dashed line}) dropouts is plotted as a function of redshift.  The probability is averaged over 5 \citet{bruzual03} SEDs discussed in the text with varying ages and metallicities and then smoothed with a boxcar average over 5 points to better represent a more diverse set of spectra.
 }
\label{fig:redshift}
\end{figure}

The completeness function gives us an estimate of the redshift distribution of our dropout samples.  Figure~\ref{fig:redshift} shows $p\left(\rm{m}_{H_{160}}=26,z\right)$ for our z$_{850}$ and J$_{110}$ dropout samples.  The completeness is also used to correct the surface density of dropouts for incompleteness.  To do this at a specific magnitude, say $\rm{m}_{H_{160}}=\rm{m}_0$, we correct the surface density using $p\left(\rm{m}_0\right)$, the probability with which we detect dropout galaxies that have the redshift distribution given by $p\left(\rm{m}_0,z\right)$.  We find $p\left(\rm{m}_{H_{160}}\right)=58\%$, $54\%$, $40\%$, and $4\%$ for $\rm{m}_{H_{160}}=24$, 25, 26, and 27, respectively.  Once again this is the average of the completeness using each of five SEDs.  As an example of how this varies with SED, the completeness at 26th magnitude ranged from 37\% to 45\% depending on the SED.
Finally, $p\left(\rm{m}_{H_{160}},z\right)$ is also used to correct the LF of dropouts as described in Section~\ref{sec:counts}.  

The ICL (occupying $\sim20-25\%$ of the search area) and obscuration by cluster members (obscuring $\sim25\%$ of the area in the regions of the search area not near the ICL) are the two main factors that lower our completeness levels.  Nevertheless, it is still advantageous to use clusters (like the Bullet Cluster) in the search for high redshift galaxies given that their magnifications facilitate reaching the more numerous faint sources overcompensating for the loss of search area obscured by the ICL.  The above completeness is an average estimate over the entire search area, including the ICL.

We have estimated the completeness as a function of lensed magnitude, $p\left(\rm{m}_{H_{160}}\right)$, above.  However, the completeness as a function of intrinsic magnitude, $p\left(\rm{m}_{H_{160,int}}\right)$, is required to estimate the intrinsic surface density.  This is the probability that a galaxy in the source plane with intrinsic magnitude $\rm{m}_{H_{160,int}}$ will be detected and satisfy our dropout color and S/N criteria.  It could be estimated by populating the source plane, properly lensing each object to the image plane, and estimating the completeness.  This is difficult however given that the objects will be multiply-imaged.  Therefore, we estimate $p\left(\rm{m}_{H_{160,int}}\right)$ as the product of the probability $p\left(\mu\right)$ that a galaxy in the source plane will be magnified by $\mu$ and the probability $p\left(\rm{m}_{H_{160}}\right)$ that if magnified by $\mu$ it will be detected as a dropout.  This product is then summed over all possible magnifications from the lens model.  This is given by
\be
p\left(\rm{m}_{H_{160,int}}\right) = \sum_{\mu} p\left(\mu\right) p\left(\rm{m}_{H_{160}}\right),
\label{eqn:compl}
\ee
where $\rm{m}_{H_{160}}=\rm{m}_{H_{160,int}}-2.5\log(\mu)$.

\begin{figure}[ht]\centering
\includegraphics[scale=0.7]{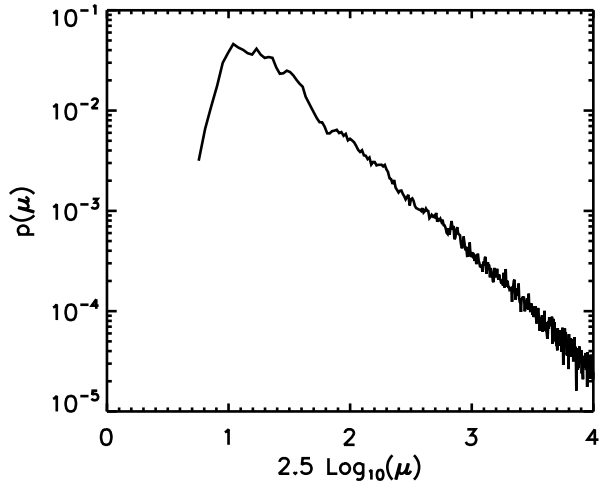}
 \caption[]{\small 
The probability, according to our lens model, that a galaxy placed randomly in the source plane will be magnified by $2.5\log(\mu)$ magnitudes.  The function is normalized such that $\sum_{\mu} p\left(\mu\right)=1$.
 }
\label{fig:pmu}
\end{figure}

We determine $p\left(\mu\right)$ (the distribution of $\mu$ values in the source plane) from the distribution of $\mu$ values in the image plane with each value weighted by the source plane area that it occupies (weighting $1/\mu$)\footnote{This does not account for multiple imaging of the source plane, which we find through simulations to be a small effect (see Section~\ref{sec:counts}).}.  This function is plotted in Figure~\ref{fig:pmu} as a function of magnification in magnitudes.  Following Equation~\ref{eqn:compl} we find $p\left(\rm{m}_{H_{160,int}}\right) =$ 56\%, 46\%, 21\%, and 5\% for $\rm{m}_{H_{160,int}}=$ 26, 27, 28, and 29, respectively. The completeness as a function of intrinsic magnitude and redshift, $p\left(\rm{m}_{H_{160,int}},z\right)$, used to correct the intrinsic LF below is also estimated following Equation~\ref{eqn:compl} with $p\left(\rm{m}_{H_{160}}\right)$ replaced by $p\left(\rm{m}_{H_{160}},z\right)$.  We have performed a check by populating the source plane and ray-tracing objects with a simple lens model to confirm that the approach of Equation~\ref{eqn:compl} is accurate.

\section{Surface Density and Luminosity Function}
\label{sec:counts}

In this section we discuss our ``lensed'' (observed) and ``intrinsic'' (true, accounting for magnification) counts in terms of the surface density and LF.  We also compare our counts with other surveys and discuss implications.  Only those objects with lensed magnitudes $\rm{m}_{H_{160}}<26.5$ where the completeness is $>22$\% are considered here.  This includes all objects in Table~\ref{tab:drops} except objects 1 and 2.

\begin{figure}[ht]\centering
\includegraphics[scale=0.7]{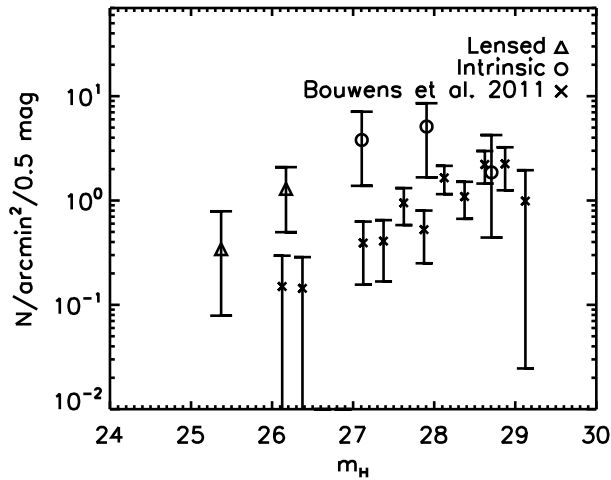}
 \caption[]{\small 
Lensed ({\it triangles}) and intrinsic ({\it circles}) completeness corrected surface density of z$_{850}$ dropouts with bins of width 0.8 magnitudes.  The lensed bins are centered at 25.4 and 26.2 and the intrinsic bins are centered at 27.1, 27.9, and 28.7.  The counts are divided by the appropriate factor for comparison with data with 0.5 magnitude bins.  The completeness corrected $z\sim7$ counts of the HUDF09 blank field survey of \citet{bouwens11} are also shown for comparison ({\it X symbols}). We do not however expect our intrinsic counts to agree with those of \citet{bouwens11} given that we use different filters and thus have different redshift distributions.  Uncertainty in our completeness estimate (see Section~\ref{sec:complete}) is not reflected in the error bars.}
\label{fig:counts}
\end{figure}

The completeness corrected surface density of our dropout sample is shown in Figure~\ref{fig:counts} for both lensed and intrinsic counts.  We first discuss the lensed surface density.  The lensed bins are 0.8 magnitudes wide and are centered at 25.4 and 26.2.  The lensed or image plane area effectively observed as a result of the completeness is given by $A_{\rm eff,len}\left(\rm{m}_{H_{160}}\right)=p\left(\rm{m}_{H_{160}}\right) \Omega_{\rm img}$, where $\Omega_{\rm img}=8.27~\rm{arcmin}^2$ is the image plane solid angle (outlined with the white dashed line in Figure~\ref{fig:bullet}).  $A_{\rm eff,len}\left(\rm{m}_{H_{160}}\right)$ was applied to each bin by dividing each dropout within the bin by $A_{\rm eff,len}\left(\rm{m}_{H_{160}}\right)$ linearly interpolated at the lensed magnitude of the dropout.

\begin{figure}[ht]\centering
\includegraphics[scale=0.7]{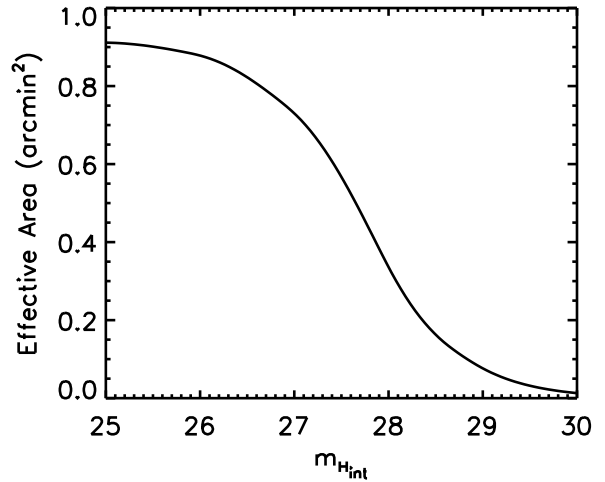}
 \caption[]{\small 
The source plane area effectively observed as a function of intrinsic H$_{160}$ magnitude.  This is the source plane area over which a galaxy with intrinsic magnitude m$_{\rm H_{int}}$ would be detected and satisfy our dropout selection criteria.  The maximum possible area at a given intrinsic magnitude assuming 100\% completeness is $\Omega_{\rm src}=1.57~\rm{arcmin}^2$.}
\label{fig:area}
\end{figure}

For the intrinsic surface density, the bins are 0.8 magnitudes wide and are centered at 27.1, 27.9, and 28.7.  The intrinsic apparent magnitude of each dropout was determined by dividing the flux by the magnification at the source position.  
The cluster lens not only magnifies the brightness of the background galaxies but also magnifies the field of view, thus decreasing the observed solid angle.  We estimate from the lens model a fractional decrease in solid angle of $0.19\pm0.02$.  The uncertainty was estimated by evaluating the solid angle change using the map of magnification errors described in Section~\ref{sec:reconstruct}.  While using the magnification maps for a different LBG redshift does change the positions of the critical curves (by $\sim5$\arcsec~ between $z=5$ and $z\rightarrow\infty$, larger than the expected accuracy of the mass model), it changes the total solid angle of the field only negligibly (within the errors given above).  The intrinsic or source plane area effectively observed as a result of the completeness (see e.g., \citealp{richard08}) is given by $A_{\rm eff,int}\left(\rm{m}_{H_{160,int}}\right)=p\left(\rm{m}_{H_{160,int}}\right) \Omega_{\rm src}$, where $\Omega_{\rm src}=1.57~\rm{arcmin}^2$ is the source plane solid angle.  This function is shown in Figure~\ref{fig:area}.  $A_{\rm eff,int}\left(\rm{m}_{H_{160,int}}\right)$ was applied to each bin by dividing each dropout within the bin by $A_{\rm eff,int}\left(\rm{m}_{H_{160,int}}\right)$ linearly interpolated at the intrinsic magnitude of the dropout.  Our determination of the intrinsic surface density and LF does not account for some of the sources being multiply-imaged (i.e., a doubly imaged source is not counted twice).  We have however verified with ray tracing, that in counting only the brightest source in the observational setup, the expected number counts change by only 15\%, which is well within the error bars as discussed next.

Errors in the number of objects in our surface density and LF include both Poisson errors and sample variance due to large scale structure, added in quadrature.  The Poisson errors\footnote{We use the sum of the fraction of the 1$\sigma$ error bars as the number of objects in each bin as discussed in Section~\ref{sec:reconstruct}.} are the 95\% confidence intervals from the Bayesian method of \citet{kraft91}.  We follow \citet{somerville04} in calculating the sample variance and assume a power-law correlation function $\xi\left(r\right)=\left({r_0/r}\right)^\gamma$.  Given the lack of constraints on the clustering of LBGs at high redshifts, we assume that the clustering of Ly$\alpha$ emitters is sufficiently similar to inform our choice of $\gamma$ and $r_0$ (see e.g., \citealp{nagamine10}).  We fix $\gamma=1.8$ and assume $r_0=7.0$ h$^{-1}$Mpc, which is conservative for $z\sim7$ Ly$\alpha$ galaxies according to \citet[Figure 5]{orsi08}.  This gives a relative sample standard deviation of $\sim38\%$ for our 1.57 arcmin$^2$ intrinsic solid angle, in agreement with the sample variance calculator of \citet{trenti08} which gives 37\%.  Given our counts, Poisson errors dominate over sample variance.

The completeness corrected $z\sim7$ counts of the HUDF09 blank field survey of \citet{bouwens11} are also shown in Figure~\ref{fig:counts}.  We expect our counts to be higher than those of \citet{bouwens11} as is the case.  This is due to the fact that \citet{bouwens11} uses z$_{850}$, Y$_{105}$ and J$_{125}$ to search for z$_{850}$ dropouts and given that Y$_{105}$ and J$_{125}$ combined cover the same wavelength range as our J$_{110}$, our redshift selection extends to higher redshift than that of \citet{bouwens11}.  We do obtain counts to only slightly shallower depths than the \citet{bouwens11} deep blank field survey with drastically less observation time, demonstrating the benefit of cluster surveys in the search for high redshift galaxies.  In addition, by searching for galaxies behind the Bullet Cluster we see galaxies magnified by factors of 2-12 (0.75-2.7 magnitudes).  

\begin{figure}[ht]\centering
\includegraphics[scale=0.7]{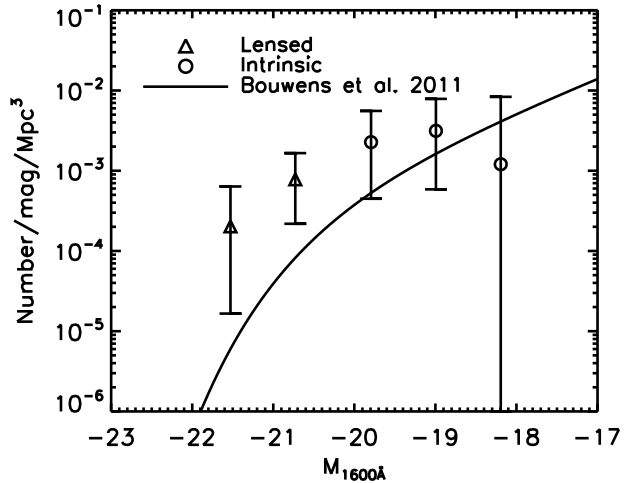}
 \caption[]{\small
Lensed ({\it triangles}) and intrinsic ({\it circles}) luminosity function of z$_{850}$ dropouts with bins of width 0.8 magnitudes.  The lensed bins are centered at  -21.5 and -20.7 and the intrinsic bins are centered at -19.8, -19, and -18.2.  The Schechter parameterization of the best-fit luminosity function of the blank field study of \citet{bouwens11} is shown for comparison ({\it solid line}, $\phi^\ast=0.9\times10^{-3}$ Mpc$^{-3}$, M$^\ast=-20.11$, and $\alpha=-1.94$).  Uncertainty in our completeness estimate (see Section~\ref{sec:complete}) is not reflected in the error bars.
 }
\label{fig:lf}
\end{figure}

The completeness corrected LF of our dropout sample is shown in Figure~\ref{fig:lf} as a function of both lensed and intrinsic magnitudes.  The best-fit $z\sim7$ LF of \citet{bouwens11} is also shown.  We first discuss the lensed LF.  The bins are 0.8 magnitudes wide and centered at -21.5 and -20.7.  The lensed LF was calculated following \citet{steidel99}.  The lensed comoving volume effectively observed as a result of the completeness is a function of lensed H$_{160}$ magnitude given by
\be
V_{\rm eff,len}\left(\rm{m}_{H_{160}}\right)=\Omega_{img} \int dz \frac{dV}{dz} p\left(\rm{m}_{H_{160}},z\right),
\label{eqn:veff}
\ee
where $dz\left(dV/dz\right)$ is the comoving volume per unit solid angle between redshift $z$ and $z+dz$ and $p\left(\rm{m}_{H_{160}},z\right)$ is the probability that an LBG of lensed magnitude $\rm{m}_{H_{160}}$ at redshift z will be detected in our images and satisfy our selection criteria (see Section~\ref{sec:complete}).  An estimate of the completeness corrected comoving number density of galaxies per magnitude bin (the LF) is then given by the lensed surface density (before dividing by $A_{\rm eff,len}$) divided by $V_{\rm eff,len}\left(\rm{m}_{H_{160}}\right)$ and the binsize.  
$V_{\rm eff,len}\left(\rm{m}_{H_{160}}\right)$ was applied to each bin by dividing each dropout within the bin by $V_{\rm eff,len}\left(\rm{m}_{H_{160}}\right)$ linearly interpolated at the lensed magnitude of the dropout.
The error bars include sample variance and Poisson errors as described above.  The fact that our lensed counts are significantly higher than the \citet{bouwens11} LF demonstrates the benefit of cluster surveys in this magnitude range as a result of the steepness of the LF leading to a positive magnification bias (see Section~\ref{sec:intro}).

We now discuss the intrinsic LF.  The bins are 0.8 magnitudes wide and centered at -19.8, -19, and -18.2.  $V_{\rm eff,int}\left(\rm{m}_{H_{160,int}}\right)$ is the intrinsic comoving volume effectively observed as a result of the completeness. It is a function of intrinsic magnitude calculated following Equation~\ref{eqn:veff} with $\Omega_{\rm img}$ replaced by $\Omega_{\rm src}$ and $p\left(\rm{m}_{H_{160}},z\right)$ replaced by $p\left(\rm{m}_{H_{160,int}},z\right)$.  The intrinsic LF is then given by the intrinsic surface density (before dividing by $A_{\rm eff,int}$) divided by $V_{\rm eff,int}\left(\rm{m}_{H_{160,int}}\right)$ and the binsize.  $V_{\rm eff,int}\left(\rm{m}_{H_{160,int}}\right)$ was applied to each bin by applying it to each dropout within the bin given its intrinsic magnitude.  

Our intrinsic LF is in agreement with that of \citet{bouwens11}.  It is not our intent to constrain the LF given so few objects, but we do provide further proof of the concept that cluster searches for high redshift LBGs are not only feasible but in many ways preferable.  These cluster surveys also provide an independent check of high redshift blank field results at much lower observing cost.  With a larger sample of massive clusters, with well understood magnification and magnification error properties (at least 2 multiply-imaged systems with known redshifts and weak-lensing data) and observed to similar depths as the Bullet Cluster, we can increase the number counts and significantly reduce the errors due to sample variance and Poisson sampling.  

\section{Conclusions}
\label{sec:conclusions}

We find 10 candidate $z\sim7$ z$_{850}$ dropout galaxies behind the Bullet Cluster.  Using the 8 objects with $\rm{m}_{H_{160}}<26.5$, where the completeness is greater than $22\%$, we calculate the surface density and luminosity function as a function of their intrinsic brightnesses.  
We find results consistent with published blank field surveys.  We thus provide an independent check of blank field results at only slightly shallower depths.
The magnifications of our z$_{850}$ dropouts range from 2-12 allowing us to detect sources up to 2.7 magnitudes deeper than blank field surveys with the same exposure time.  With the magnification of the Bullet Cluster we are thus able to probe to similar depths as blank field surveys despite using much shallower data.

In searching for high redshift LBGs, we find 8 (3) objects with $\rm{m}_{H_{160,{\rm int}}}<25$ that although they do not satisfy some of our z$_{850}$ (J$_{110}$) dropout selection criteria and are thus rejected as low redshift contaminants, they are sufficiently red to satisfy our color cuts and are thus interesting in their own right.  We give their positions and photometric properties.  

Magnifications are calculated from an optimally combined weak and strong lensing mass reconstruction of the Bullet Cluster \citep{bradac09}.  Errors on the magnification are smaller than the Poisson sampling and sample variance.  With more clusters with mass models of similar quality as the Bullet Cluster we could significantly reduce these errors and probe the luminosity function at greater depths and higher redshifts than is otherwise possible.  With a larger sample of efficient lenses we would also expect to find many highly magnified images, which would allow for better morphological studies and greatly ease the spectroscopic follow up of $z\gtrsim7$ galaxies. 

\acknowledgments

We thank Rychard Bouwens for helpful discussions as well as providing relevant data for comparison with our results.  We also thank Adriano Fontana and Laura Pentericci for valuable discussions.
Support for this work was provided by NASA through grant numbers HST-GO-10200, HST-GO-10863, and HST-GO-11099 from the Space Telescope Science Institute (STScI), which is operated by AURA, Inc., under NASA contract NAS 5-26555 and NNX08AD79G. Tommaso Treu acknowledges support from the NSF through CAREER award NSF-0642621, by the Sloan Foundation through a Sloan Research Fellowship, and by the Packard Foundation through a Packard Fellowship.

\bibliography{marusa}
\end{document}